# Spectroscopic Evidence for Competing Order-Induced Pseudogap Phenomena and Unconventional Low-Energy Excitations in High-$T_C$ Cuprate Superconductors

N.-C. Yeh*, A. D. Beyer, M. L. Teague, S.-P. Lee*, S. Tajima†, S. I. Lee‡

**Abstract** The low-energy excitations of cuprate superconductors exhibit various characteristics that differ from those of simple Bogoliubov quasiparticles for pure $d_{x^2-y^2}$-wave superconductors. Here we report experimental studies of spatially resolved quasiparticle tunnelling spectra of hole- and electron-type cuprate superconductors that manifest direct evidences for the presence of competing orders (COs) in the cuprates. In contrast to conventional type-II superconductors that exhibit enhanced local density of states (LDOS) peaking at zero energy near the centre of field-induced vortices, the vortex-state LDOS of $YBa_2Cu_3O_{7-\delta}$ (Y-123) and $La_{0.1}Sr_{0.9}CuO_2$ (La-112) remains suppressed inside the vortex core, with pseudogap (PG)-like features at an energy larger (smaller) than the superconducting (SC) gap $\Delta_{SC}$ in Y-123 (La-112). Energy histograms of the SC and PG features reveal steady spectral shifts from SC to PG with increasing magnetic field $H$. These findings may be explained by coexisting COs and SC: For hole-type cuprates with PG above $T_c$, the primary CO gap ($V_{CO}$) is *larger* than $\Delta_{SC}$ and the corresponding COs are charge/pair-density waves with wave-vectors parallel to $(\pi,0)/(0,\pi)$. For electron-type cuprates without PG above $T_c$, $V_{CO}$ is *smaller* than $\Delta_{SC}$ and the CO wave-vector is along $(\pi,\pi)$. This CO scenario may be extended to the ARPES data to consistently account for the presence (absence) of Fermi arcs in hole- (electron)-type cuprates. Fourier transformation of the vortex-state LDOS in Y-123 further reveals multiple sets of energy-independent wave-vectors due to field-enhanced pair- and spin-density waves. These results imply important interplay of SC with low-energy collective excitations.

## 1 Introduction: Competing Orders in Doped Mott Insulators

High-temperature superconducting cuprates are doped Mott insulators with strongly correlated electronic ground states [1]. The complexity of these materials and the strong electronic correlation gives rise to various competing orders (COs) in the ground state besides superconductivity (SC), as manifested by such experimental evidences as x-ray and neutron scattering, $\mu$SR, NMR, Raman scattering, ARPES, STM [2-15] and further confirmed by theoretical modeling/simulations [16-25]. The occurrence of specific types of COs such as the charge-density waves (CDW) [16,17], pair-density waves (PDW) [21,22], or spin-density waves (SDW) [18-20] depends on the microscopic properties of a given cuprate, which include: electron or hole-doping, doping level ($\delta$), number of $CuO_2$ layers per unit cell ($n$), and electronic anisotropy ($\gamma$). Although the relevance of COs to cuprate SC remains unclear, the existence of COs has a number of important physical consequences. First, quantum criticality naturally emerges as the result of competing phases in the ground state [5,18,25,26]. Second, strong quantum fluctuations are expected as the result of proximity to quantum criticality [5,14,27]. Third, the low-energy excitations are unconventional as the result of redistributions of the spectral weight between SC and COs [11-15,28-30]. The unconventional phenomena include: satellite features [3,5,14,28-30] and periodic LDOS modulations in the quasiparticle spectra of hole-type cuprates [11-13,30]; the excess sub-gap DOS in electron-type cuprates below $T_c$ [5,14]; "dichotomy" in the momentum dependence of quasiparticle coherence [2,28,31,32]; and PG-like vortex-core states [15,29,30]. Fourth, the exact CO phase and the energy scale may differ amongst the cuprates, leading to various non-universal phenomena [5,14] such as the presence or absence of the low-energy PG [2,3,28,29], anomalous Nernst effect [33] and Fermi arcs [2,10,34]; the varying spatial homogeneity and modulations in the quasiparticle spectra [15,30,31,35-37]; and the characteristics of magnetic excitations [38-40]. Finally, The presence of COs and strong quantum fluctuations naturally lead to weakened superconducting stiffness upon increasing $T$ and $H$ [5,27,41,42] and may be responsible for the extreme type-II nature [43]. Hence, COs may be highly relevant to the strong fluctuations and novel vortex dynamics in cuprate superconductors [5,27].

*Department of Physics, California Institute of Technology, Pasadena, CA 91125, USA. E-mail: ncyeh@caltech.edu
†Department of Physics, Osaka University, Osaka 560-0043, Japan.
‡Department of Physics, Sogang University, Seoul, Korea 121-742.

For comparison, an alternative theoretical viewpoint commonly referred to as the "preformed pair" model or the "one-gap" model assumes that the low-energy PG temperature $T^*$ is the onset of Cooper pairing and the superconducting transition $T_c$ is the onset of phase coherence [41,44-47]. However, the phenomenology associated with the one-gap scenario is only partially applicable to the hole-type cuprates. Additionally, the one-gap notion cannot account for either the appearance of energy-independent wave-vectors or charge modulations that are doping-dependent in the SC state of hole-type cuprates [13]. In contrast, the CO scenario, or the "two-gap" model, is not exclusive of the possibility of preformed pairs: COs represent additional phase instabilities in the cuprates that are neglected in all one-gap models, and they may coexist with preformed pairs above $T_c$ if they already coexist with coherent Cooper pairs at low temperatures.

In this work, we describe our scanning tunneling spectroscopic (STS) studies of quasiparticle spectra of various cuprates as functions of temperature ($T$) and magnetic field ($H$). We also present our own theoretical analyses based on the CO scenario for the quasiparticle tunneling spectra and ARPES data of both electron- and hole-type cuprates with varying doping levels, and find that a unified phenomenology emerges. These studies therefore suggest an important interplay between collective low-energy bosonic excitations and cuprate superconductivity.

## 2 Scanning Tunneling Spectroscopic Studies of the Low-Energy Quasiparticle Excitations

The spatially resolved tunneling conductance ($dI/dV$) versus energy ($\omega = eV$) spectra for the quasiparticle LDOS maps were obtained with our homemade cryogenic scanning tunneling microscope (STM). Our STM has a base temperature of 6 K, variable temperature range up to room temperature, magnetic field range up to 7 Tesla, and ultra-high vacuum capability down to a base pressure < $10^{-9}$ Torr at 6 K. For each constant temperature ($T$) and magnetic field ($H$), the experiments were conducted by tunneling currents along the crystalline c-axis under a range of bias voltages at a given location. The typical junction resistance was $\sim 1$ G$\Omega$. Current ($I$) vs. voltage ($V$) measurements were repeated pixel-by-pixel over an extended area of the sample. To remove slight variations in the tunnel junction resistance from pixel to pixel, the differential conductance at each pixel is normalized to its high-energy background. More details of our experimental setup, surface preparation and tunneling conditions have been described elsewhere [15,30,35]. The cuprates in our spatially resolved tunneling spectra study include optimally doped hole-type YBa$_2$Cu$_3$O$_{7-\delta}$ (Y-123) with $T_c$ = 93 K and optimally doped electron-type La$_{0.1}$Sr$_{0.9}$CuO$_2$ (La-112) with $T_c$ = 43 K. For analysis, we apply Green function techniques based on the CO scenario [28,29] to our tunneling spectra as well as to spectra taken by others on such systems as hole-type Bi$_2$Sr$_2$CaCu$_2$O$_{8+x}$ (Bi-2212) [31] and Bi$_2$Sr$_2$CuO$_{6+x}$ (Bi-2201) [13] and electron-type Pr$_{1.85}$Ce$_{0.15}$CuO$_{4-x}$ (PCCO) [48] of various doping levels in zero-field for self-consistent comparison.

2.1 Temperature-Dependent Tunneling Spectra in Zero Fields

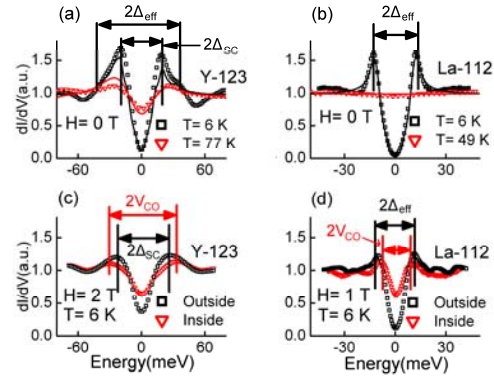

**Fig. 1** Implication of CO from zero- and finite-field STS in Y-123 and La-112: **(a)** Normalized zero-field tunneling spectra of Y-123 taken at $T$ = 6 K (black) and 77 K (red). The solid lines represent fittings to the $T$ = 6 and 77 K spectra by assuming coexisting SC and CDW, with fitting parameters of $\Delta_{SC}$ = 20 meV, $V_{CDW}$ = 32 meV and Q$_{CDW}$ = (0.25$\pi$ ± 0.05$\pi$, 0) / (0, 0.25$\pi$ ± 0.05$\pi$), following Refs. [28,29]. **(b)** Normalized zero-field tunneling spectra of La-112 taken at $T$ = 6 K (black) and 49 K (red). The solid lines represent fittings to the $T$ = 6 and 49 K spectra by assuming coexisting SC and SDW, with fitting parameters $\Delta_{SC}$ = 12 meV, V$_{SDW}$ = 8 meV, and Q$_{SDW}$ = (±$\pi$, ±$\pi$), following Refs. [28,29]. **(c)** Spatially averaged intra- and inter-vortex spectra of Y-123 for $T$ = 6 K and $H$ = 2 T, showing PG features inside the vortex, with a PG energy *larger* than $\Delta_{SC}$ and consistent with the $V_{CDW}$ value derived from fitting the zero-field spectra in (a). **(d)** Spatially averaged intra- and inter-vortex spectra of La-112 for $T$ = 6 K and $H$ = 1 T, showing PG features inside the vortex, with a PG energy *smaller* than $\Delta_{SC}$ and consistent with the $V_{SDW}$ value derived from fitting the zero-field spectra in (b).

In Figs. 1(a) and 1(b) the *T*-evolution of the quasiparticle tunneling spectra are exemplified for optimally doped cuprate superconductors Y-123 and La-112, together with the corresponding theoretical fittings based on the CO scenario and realistic bandstructures. The Green function techniques have been detailed by us in Refs. [28,29] and are briefly outlined in the following.

Our theoretical analysis begins with a mean-field Hamiltonian $\mathcal{H}_{MF} = \mathcal{H}_{SC} + \mathcal{H}_{CO}$ that consists of coexisting SC and a CO at $T = 0$ [28,29]. We further assume that the SC gap $\Delta_{SC}$ vanishes at $T_c$ and the CO order parameter vanishes at $T^*$, and that both $T_c$ and $T^*$ are second-order phase transitions. The SC Hamiltonian is given by:

$$\mathcal{H}_{SC} = \sum_{\mathbf{k},\alpha} \xi_{\mathbf{k}} c^\dagger_{\mathbf{k},\alpha} c_{\mathbf{k},\alpha} - \sum_{\mathbf{k}} \Delta_{SC}(\mathbf{k}) \left( c^\dagger_{\mathbf{k},\uparrow} c^\dagger_{-\mathbf{k},\downarrow} + c_{-\mathbf{k},\downarrow} c_{\mathbf{k},\uparrow} \right), \quad (1)$$

where $\Delta_{SC}(\mathbf{k}) = \Delta_{SC}(\cos k_x - \cos k_y)/2$ for $d_{x^2-y^2}$-wave pairing, $\mathbf{k}$ denotes the quasiparticle momentum, $\xi_{\mathbf{k}}$ is the normal-state eigen-energy relative to the Fermi energy, $c^\dagger$ and $c$ are the creation and annihilation operators, and $\alpha = \uparrow, \downarrow$ refers to the spin states. The CO Hamiltonian is specified by the energy $V_{CO}$, a wave-vector $\mathbf{Q}$, and a momentum distribution $\delta\mathbf{Q}$ that depends on a form factor, the correlation length of the CO, and also on the degree of disorder. We have previously considered the effect of various types of COs on the quasiparticle spectral density function $A(\mathbf{k},\omega)$ and the density of states $\mathcal{N}(\omega)$. For instance, in the case that charge density waves (CDW) is the relevant CO, we have a wave-vector $\mathbf{Q}_1$ parallel to the CuO$_2$ bonding direction $(\pi,0)$ or $(0,\pi)$ in the CO Hamiltonian [28,29]:

$$\mathcal{H}_{CDW} = -\sum_{\mathbf{k},\alpha} V_{CDW}(\mathbf{k}) \left( c^\dagger_{\mathbf{k},\alpha} c_{\mathbf{k}+\mathbf{Q}_1,\alpha} + c^\dagger_{\mathbf{k}+\mathbf{Q}_1,\alpha} c_{\mathbf{k},\alpha} \right). \quad (2)$$

On the other hand, for commensurate SDW being the relevant CO, the SDW wave-vector becomes $\mathbf{Q}_2 = (\pi,\pi)$, and the corresponding CO Hamiltonian is [49]:

$$\mathcal{H}_{SDW} = -\sum_{\mathbf{k},\alpha,\beta} V_{SDW}(\mathbf{k}) \left( c^\dagger_{\mathbf{k}+\mathbf{Q}_2,\alpha} \sigma^3_{\alpha\beta} c_{\mathbf{k},\beta} + c^\dagger_{\mathbf{k},\alpha} \sigma^3_{\alpha\beta} c_{\mathbf{k}+\mathbf{Q}_2,\beta} \right) \quad (3)$$

where $\sigma^3_{\alpha\beta}$ denotes the matrix element $\alpha\beta$ of the Paul matrix $\sigma^3$.

Thus, by incorporating realistic bandstructures and Fermi energies for different families of cuprates with given doping and by specifying the SC pairing symmetry and the form factor for the CO, we can diagonalize $\mathcal{H}_{MF}$ to obtain the bare Green function $G_0(\mathbf{k},\omega)$ for momentum $\mathbf{k}$ and energy $\omega$. We may further include quantum phase fluctuations between the CO and SC and then solve the Dyson's equation self-consistently for the full Green function $G(\mathbf{k},\omega)$ [28,29], which gives the quasiparticle spectral density function $A(\mathbf{k},\omega) = -\text{Im}[G(\mathbf{k},\omega)]/\pi$ for comparison with ARPES [34] and the quasiparticle density of states $\mathcal{N}(\omega) = \sum_{\mathbf{k}} A(\mathbf{k},\omega)$ for comparison with STM spectroscopy [28,29].

Based on the Green function analysis outlined above for coexisting $d_{x^2-y^2}$-wave SC and a specific CO, the zero-field quasiparticle spectra $\mathcal{N}(\omega)$ and $A(\mathbf{k},\omega)$ at $T = 0$ can be fully determined by the parameters $\Delta_{SC}$, $V_{CO}$, $\mathbf{Q}$, $\delta\mathbf{Q}$, $\Gamma_{\mathbf{k}}$ (the quasiparticle linewidth), and $\eta$ (the magnitude of quantum phase fluctuations), which is proportional to the mean-value of the velocity-velocity correlation function [28,29]. For finite temperatures, we employ the temperature Green function.

Using the aforementioned theoretical analysis we have been able to consistently account for the *T*-dependent quasiparticle tunneling spectra in both hole- and electron-type cuprates if we assume Fermi-surface nested CDW [28,29] as the CO in the hole-type cuprates such as Y-123 and Bi-2212, and commensurate SDW as the CO in the electron-type La-112 and PCCO, which are consistent with findings from neutron scattering experiments [40]. Specifically, for hole-type cuprates such as in the spectra of Y-123, the sharp peaks and satellite "hump" features at $T \ll T_c$ in Fig. 1(a) are associated with $\omega = \pm\Delta_{SC}$ and $\omega = \pm\Delta_{eff}$, respectively, where $\Delta_{eff} \equiv [(\Delta_{SC})^2 + (V_{CO})^2]^{1/2}$ is an effective excitation gap. Hence, the condition $V_{CO} > \Delta_{SC}$ in hole-type cuprates is responsible for the appearance of the satellite features at $T \ll T_c$ and the PG phenomena at $T^* > T > T_c$ [28,29,34]. In contrast, the condition $V_{CO} < \Delta_{SC}$ in electron-type cuprates, as exemplified in Fig. 1(b), is responsible for only one set of characteristic features at $\omega = \pm\Delta_{eff}$ and the absence of PG above $T_c$.

We have extended our analysis to tunneling spectra of different doping levels ($\delta$) associated with hole-type cuprates [29]. We find that $\Delta_{SC}(\delta)$ generally follows the same non-monotonic dependence of $T_c(\delta)$. In contrast, $V_{CO}(\delta)$ increases with decreasing $\delta$, which is consistent with the general trend of the zero-field PG temperature in hole-type cuprates [29].

### 2.2 Spatially Resolved Vortex-State Quasiparticle Tunneling Spectra in Y-123 and La-112

An alternative way of verifying the feasibility of the CO scenario is to introduce vortices because the suppression of SC inside vortices may unravel the spectroscopic characteristics of the remaining CO. As exemplified in Fig. 1(c) for a set of intra- and inter-vortex spectra taken on Y-123 at $H = 2$ T and 6 K and in Fig. 1(d) for a set of intra- and inter-vortex spectra taken on La-112 at $H = 1$ T, the quasiparticle spectra near the center of each vortex exhibit pseudogap (PG)-like features, which is in stark contrast to theoretical predictions for a sharp zero-energy peak around the center of the vortex core had SC been the sole order in the ground state [50-52]. Interestingly, for both Y-123 and La-112, the respective PG energy inside vortices is comparable to the CO energy $V_{CO}$ derived from our zero-field fittings in Figs. 1(a) and 1(b). Additionally, a subgap feature at $\Delta' < \Delta_{SC}$ is found inside the vortex cores of Y-123.

To investigate how quasiparticle spectra evolve with magnetic field, we performed spatially resolved spectroscopic studies at varying fields for both Y-123 and La-112 at $T = 6$ K. In Figs. 2(b) and 2(d) we show exemplified spatial maps of the conductance power ratio $r_G$ at $H = 2$ T for Y-123 and La-112, respectively. Here $r_G$ at every pixel is defined as the ratio of the conductance power $(dI/dV)^2$ at $|\omega| = \Delta_{SC}$ relative to that at $\omega = 0$. We find that the presence of vortices is associated with the local minimum of $r_G$ because of enhanced zero-energy quasiparticle density of states inside the vortex core. Moreover, the total flux is conserved within the area studied despite the appearance of disordered vortices. That is, the total number of vortices multiplied by the flux quantum is equal to the magnetic induction multiplied by the area, within experimental errors. The average radius of the vortices in Y-123 is significantly larger than the superconducting coherence length, consistent with those reported for Bi-2212 [53] and $Na_xCa_{2-x}CuO_2Cl_2$ (Na-CCOC) [54]. In contrast, for La-112 the vortex radius is comparable to the SC coherence length [15].

For comparison, we also show spatial maps of the conductance power ratio $r_G$ at $H = 0$ for Y-123 in Fig. 2(a) and for La-112 in Fig. 2(c). Clearly the $r_G$ maps at $H = 0$ for both samples are much more homogeneous than those at finite fields, which confirm our observation of vortices [15,30].

To attain further insights into the vortex-state quasiparticle spectra, we consider the spatially resolved tunneling conductance as a function of energy, which provides information for the energy evolution of the quasiparticle local density of states (LDOS). As exemplified in Figs. 3(a) – 3(d) for the LDOS of Y-123 under $H = 5$ T at various energies over a $(22 \times 29)$ nm$^2$ area, we find that the vortex-state LDOS exhibits dominating density-wave like modulations in addition to the vortices. Moreover, the inter-vortex LDOS modulations are nearly energy-independent, whereas the intra-vortex LDOS modulations exhibit energy-dependent contrasts, being maximum at $|\omega| <\sim \Delta_{SC}$ and nearly vanishing for $|\omega| >> V_{CO}$.

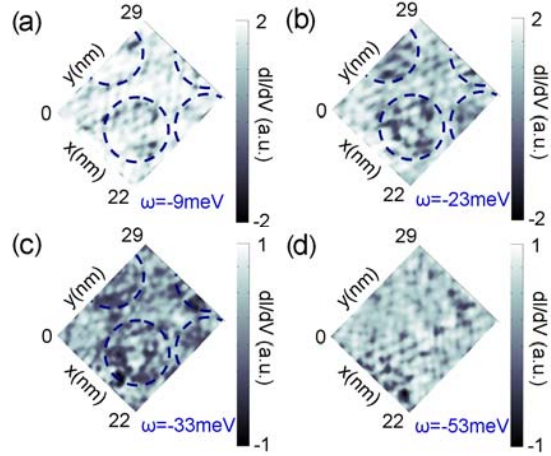

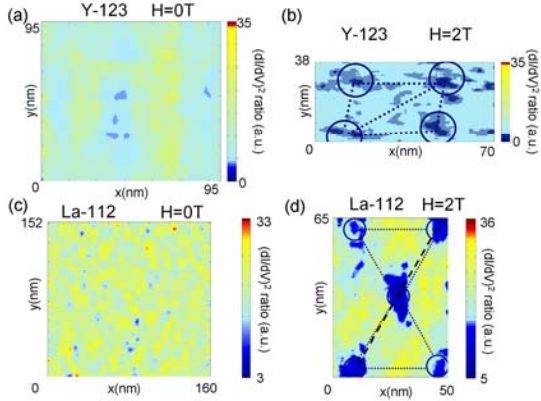

**Fig. 2** Spatial maps of the conductance power ratio $r_G$ for Y-123 and La-112 at $T = 6$ K and for $H \parallel$ c-axis: **(a)** The $r_G$ map of Y-123 taken at $H = 0$ over a $(95 \times 95)$ nm$^2$ area, showing a relatively long-range homogeneity. **(b)** The $r_G$ map of Y-123 taken over a $(70 \times 38)$ nm$^2$ area for $H = 2$ T, showing disordered vortices with an average vortex lattice constant $a_B \sim 33$ nm. **(c)** The $r_G$ map of La-112 taken at $H = 0$ over a $(160 \times 152)$ nm$^2$ area, showing long-range homogeneity relative to that in (d). Here $r_G$ is defined as the ratio of $(dI/dV)^2$ at $|\omega| = \Delta_{eff}$ and that at $\omega = 0$. **(d)** The $r_G$ map of La-112 taken at $H = 2$ T over a $(50 \times 65)$ nm$^2$ area, showing an averaged $a_B \sim 35$ nm.

**Fig. 3** The LDOS modulations of Y-123 at $H = 5$T and 6 K over a $(22 \times 29)$ nm$^2$ area, showing patterns associated with density-wave modulations and vortices (circled objects) for $\omega =$ (a) −9 meV ~ −$\Delta'$, (b) −23 meV ~ −$\Delta_{SC}$, (c) −33 meV ~ −$V_{CO}$ and (d) −53 meV. We note that the vortex contrasts are the most apparent at $|\omega| <\sim \Delta_{SC}$ and become nearly invisible for $|\omega| >> V_{CO}$. The vanishing contrast at high energies may be due to the onset of Cu-O optical phonons (~ 50 meV for the cuprates [55]) so that both the collective modes and quasiparticles become scattered inelastically.

To better understand the LDOS modulations, we perform Fourier transformation (FT) of the spatially resolved vortex-state LDOS data. Systematic analysis of the energy dependence of the FT-LDOS, $F(\mathbf{k},\omega)$, reveals two types of diffraction spots in the momentum space [30]. One type of spots are strongly energy dependent and may be attributed to elastic quasiparticle scattering interferences as seen in the zero-field FT-LDOS [31,56]. The other type of spots are nearly energy-*independent*, which are circled in Figs. 4(a) and 4(b) for FT-LDOS at $\omega = -12$ meV and for $H = 0$ and 5 T, respectively. In addition to the reciprocal lattice vectors and the $(\pi,\pi)$ resonance mode [38], we find two sets of nearly energy-independent wave-vectors along $(\pi,0)/(0,\pi)$, which are denoted by $\mathbf{Q}_{PDW}$ and $\mathbf{Q}_{CDW}$; and one set of energy-independent wave-vector along $(\pi,\pi)$, which is denoted by $\mathbf{Q}_{SDW}$. Quantitatively, $\mathbf{Q}_{PDW} = [\pm(0.56 \pm 0.06)\pi/a_1, 0]$ and $[0, \pm(0.56 \pm 0.06)\pi/a_2]$, $\mathbf{Q}_{CDW} = [\pm(0.28 \pm 0.02)\pi/a_1, 0]$ and $[0, \pm(0.28 \pm 0.02)\pi/a_2]$, and $\mathbf{Q}_{SDW} = [\pm(0.15 \pm 0.01)a_1, \pm(0.15 \pm 0.01)a_2]$. Here $a_1 = 0.383$ nm and $a_2 = 0.388$ nm.

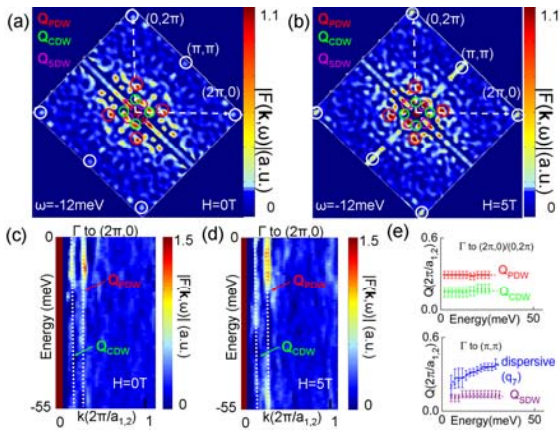

**Fig. 4** FT studies of the conductance maps of Y-123: **(a)** The intensity of the FT-LDOS, $|F(\mathbf{k},\omega)|$, is shown in two-dimensional momentum space for $H = 0$ and $\omega = -12$ meV. **(b)** $|F(\mathbf{k},\omega)|$ at $H = 5$T and $\omega = -12$ meV. Comparing (a) and (b) as well as the evolution of $|F(\mathbf{k},\omega)|$ with $\omega$, we identify three sets of $\omega$-independent wave-vectors in addition to the reciprocal lattice constants and the $(\pi,\pi)$ resonance: $\mathbf{Q}_{PDW}$ and $\mathbf{Q}_{CDW}$ along $(\pi,0)/(0,\pi)$ and $\mathbf{Q}_{SDW}$ along $(\pi,\pi)$, which are circled for clarity. **(c)** The $\omega$-dependence of $|F(\mathbf{k},\omega)|$ at $H = 0$ is plotted against $\mathbf{k} \parallel (\pi,0)$, showing $\omega$-independent modes (bright vertical lines) at $\mathbf{Q}_{PDW}$ and $\mathbf{Q}_{CDW}$. **(d)** The $\omega$-dependence of $|F(\mathbf{k},\omega)|$ at $H = 5$ T is plotted against $\mathbf{k} \parallel (\pi,0)$, showing field-enhanced intensity at $\mathbf{k} = \mathbf{Q}_{PDW}$ and $\mathbf{Q}_{CDW}$. **(e)** Upper panel: Momentum ($|\mathbf{q}|$) vs. energy ($\omega$) for $\mathbf{q} = \mathbf{Q}_{PDW}$ and $\mathbf{Q}_{CDW}$ along $(\pi,0)$. Lower panel: $|\mathbf{q}|$ vs. $\omega$ for $\mathbf{q} = \mathbf{Q}_{SDW}$ along $(\pi,\pi)$. A representative dispersive wave-vector $\mathbf{q}_7$ along $(\pi,\pi)$ due to quasiparticle scattering interferences [31,56] is also shown in the lower panel for comparison.

For clarity, we illustrate in Figs. 4(c) and 4(d) the intensity of $|F(\mathbf{k},\omega)|$ on the $|\mathbf{k}|$-vs.-$\omega$ plane for $\mathbf{k} \parallel (\pi,0)$ at $H = 0$ and 5 T, respectively. We find that strong intensities occur as vertical lines at $|\mathbf{k}| = |\mathbf{Q}_{CDW}|$ and $|\mathbf{Q}_{PDW}|$. Furthermore, the intensity of $|F(\mathbf{k},\omega)|$ at $|\mathbf{k}| = |\mathbf{Q}_{CDW}|$ and $|\mathbf{Q}_{PDW}|$ are enhanced by magnetic fields. As an alternative illustration, we plot in the upper panel of Fig. 4(e) the $\omega$-dependence of $\mathbf{Q}_{CDW}$ and $\mathbf{Q}_{PDW}$ for $\mathbf{k} \parallel (\pi,0)$, and in the lower panel the $\omega$-dependence of $\mathbf{Q}_{SDW}$ for $\mathbf{k} \parallel (\pi,\pi)$. Apparently $\mathbf{Q}_{CDW}$, $\mathbf{Q}_{PDW}$ and $\mathbf{Q}_{SDW}$ are all nearly $\omega$-independent. For comparison, a dispersive wave-vector associated with quasiparticle scattering interferences [31,56] along $(\pi,\pi)$, which is denoted as $\mathbf{q}_7$ in Ref. [31], is also shown in the lower panel of Fig. 4(e). We find that the dispersion relation for the mode $\mathbf{q}_7$ along the nodal direction is in good agreement with both the experimental results found in Bi-2212 [31] and the theoretical predictions for quasiparticle scattering interferences [31,56].

To elucidate the nature of these $\omega$-independent wave-vectors for LDOS modulations, we consider the symmetry of the complex quantity Re[$F(\mathbf{k},\omega)$] relative to $\omega$ at $\mathbf{k} = \mathbf{Q}_{XDW}$ (X = C, P, S) in Figs. 5(a) – 5(d), with the symmetric and anti-symmetric components of Re[$F(\mathbf{k},\omega)$] at $\mathbf{k} = \mathbf{Q}_{XDW}$ shown in the upper and lower panels, respectively. We find that the energy dependences of Re[$F(\mathbf{k},\omega)$] at $\mathbf{Q}_{PDW}$ and $\mathbf{Q}_{CDW}$ are similar to theoretical predictions [21,22], although understanding for the quantitative details requires further studies. We further note that the $\mathbf{Q}_{PDW}$ mode found in our FT-LDOS in fact agrees with the checkerboard-like modulations reported previously for Bi-2212 [11,12] and Bi-2201 [13]. Moreover, the empirically derived $\mathbf{Q}_{CDW}$ is consistent with the theoretical value of the CO wave-vector derived from our Green function analysis for the zero-field tunneling spectra. We therefore assign the two $\omega$-independent modes along $(\pi,0)/(0,\pi)$ to PDW and CDW. On the other hand, Re[$F(\mathbf{k},\omega)$] for $\mathbf{Q}_{SDW}$ along $(\pi,\pi)$ appears to have dominating anti-symmetric components in finite fields, and is tentatively attributed to the SDW for particle-hole excitations in the $(\pi,\pi)$ direction. However, we note that there have not been theoretical calculations for the Re[$F(\mathbf{k},\omega)$] components of SDW for comparison.

Finally, in Figs. 6(a) and 6(b) we illustrate the histograms of the vortex-state characteristic energies in Y-123 and La-112, respectively, as functions of magnetic field. In the case of Y-123, a strong spectral shift from SC at $\omega = \Delta_{SC}$ to PG at $\omega = V_{CO} > \Delta_{SC}$ is seen with increasing $H$, together with the appearance of a third subgap (SG) feature at $\omega = \Delta' < \Delta_{SC}$. In contrast, for La-112 the energy histogram at each magnetic field can be fit by a Lorentzian functional form with a peak energy at $\Delta_{eff}(H)$, which decreases slightly with increasing $H$. Additionally, there is an apparent low energy "cutoff" at $V_{CO}$ ($\sim 8$ meV) $< \Delta_{SC}$ ($\sim 12$ meV) for all energy histograms. Neither the histograms of Y-123 nor those of La-112 exhibit any spectral peak at $\omega = 0$, which are in sharp contrast to the steady increase of the spectral weight at $\omega = 0$ for conventional type-II superconductors. Therefore, we have shown that the field-revealed energy gaps $\Delta'$ and $V_{CO}$ and the $\omega$-independent wave-vectors for the LDOS modulations in the cuprates cannot be explained by assuming pure SC in the ground state.

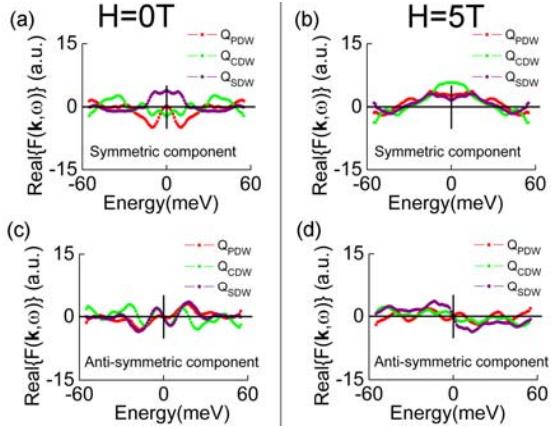

**Fig. 5** The symmetric and anti-symmetric components of Re[$F(\mathbf{k},\omega)$] for $\mathbf{k} = \mathbf{Q}_{PDW}$, $\mathbf{Q}_{CDW}$ and $\mathbf{Q}_{SDW}$ as functions of $\omega$ and $H$: **(a)** The symmetric components of Re[$F(\mathbf{k},\omega)$] at $H = 0$; **(b)** The symmetric components of Re[$F(\mathbf{k},\omega)$] at $H = 5$ T; **(c)** The anti-symmetric components of Re[$F(\mathbf{k},\omega)$] at $H = 0$; **(d)** The anti-symmetric components of Re[$F(\mathbf{k},\omega)$] at $H = 5$ T.

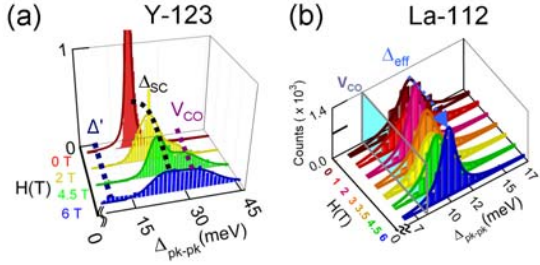

**Fig. 6** Magnetic field-dependent spectral evolution in cuprate superconductors at $T = 6$ K: **(a)** Energy histograms derived from the quasiparticle tunneling spectra of Y-123 for $H = 0$, 2, 4.5, and 6 T, showing a steady spectral shift from $\Delta_{SC}$ to $V_{CO}$ and $\Delta'$ with increasing $H$. **(b)** Magnetic field dependence of the characteristic energies $\Delta_{eff}$, $\Delta_{SC}$ and $V_{CO}$ in La-112 for $H = 0$, 1, 2, 3, 3.5, 4.5 and 6 T. Each histogram can be fit with a Lorentzian functional form. The peak position of the Lorentian is identified as $\Delta_{eff}(H)$, and the low-energy cutoff of the histogram is identified as $V_{CO}$. Empirically, we find that $V_{CO}$ is nearly constant, whereas $\Delta_{eff}(H)$ decreases slightly with increasing $H$.

### 3 Application of the Competing Order Scenario to ARPES Data

In addition to the effect of COs on the quasiparticle tunneling spectra of hole- and electron-type cuprates, we have investigated how COs may influence the angle-resolved photoemission spectroscopy (ARPES) [34]. We find that the appearance of the low-energy PG in hole-type cuprates may be correlated with the appearance of the "Fermi arc" above $T_c$ and below the PG temperature $T^*$ [2,10,34] within the CO scenario. Here the Fermi arc refers to the truncated Fermi surface not fully recovered at $T_c < T < T^*$ [2,10,34]. As exemplified in Fig. 7(a) for a slightly underdoped Bi-2212 and detailed elsewhere [34], we have shown that the Fermi arc as a function of the quasiparticle momentum $\mathbf{k}$, temperature ($T$) and doping level ($\delta$) in Bi-2212 [10] may be explained consistently by assuming incommensurate CDW with $V_{CDW} > \Delta_{SC}$ as the relevant CO. Similarly, the $\mathbf{k}$- and $T$-dependence of the effective gap $\Delta_{eff}$ and the absence of Fermi arcs in electron-type cuprates (*e.g.* $Pr_{0.89}LaCe_{0.11}CuO_4$, denoted as PLCCO) [56] can also be explained by incorporating commensurate SDW with $V_{SDW} < \Delta_{SC}$ into spectral characteristics [34]. Hence, we have shown that the CO scenario can provide unified phenomenology for a wide variety of experimental findings.

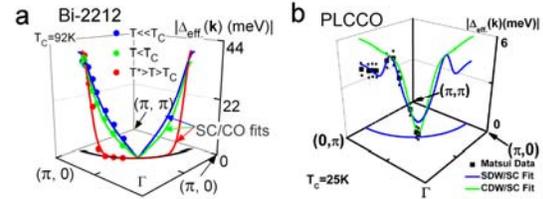

**Fig. 7** Comparison of the $\Delta_{eff}(\mathbf{k})$ vs. $\mathbf{k}$ data (symbols) and theoretical fittings (lines) between a hole-type cuprate (Bi-2212) and an electron-type cuprate (PLCCO): **(a)** $\Delta_{eff}(\mathbf{k})$ of a slightly underdoped Bi-2212 with $T_c = 92$ K, $T^* = 150$ K, $\delta = 0.15$ [10,34]. The fitting parameters $\Delta_{SC}(T)$ and $V_{CO}(T)$ for $T = 10, 82, 102$ K are given in Ref. [34]. Additionally, $|\mathbf{Q}| = 0.2\pi$ and $|\delta\mathbf{Q}| = 0.18\pi, 0.17\pi, 0.1\pi$ for $T = 10, 82, 102$ K. **(b)** Momentum dependent ARPES leading edge data (×2) from Ref. [57] are shown as a function of $\phi \equiv \tan^{-1}(k_y/k_x)$, together with theoretical fittings for two types of COs, CDW and SDW. The navy (dark) line corresponds to $\mathbf{Q} = (\pi,\pi)$ for SDW, and the green (light) line corresponds to $\mathbf{Q} \parallel (\pi,0)/(0,\pi)$ for CDW. Clearly the fitting curve with $\mathbf{Q} = (\pi,\pi)$ for SDW agrees much better with ARPES data. Moreover, the presence of commensurate SDW is also consistent with the findings of neutron scattering data on one-layer electron-type cuprates [40].

### 4 Competing Orders as the Physical Origin for Spatially Inhomogeneous LDOS in Bi-2212

In the context of spatial homogeneity of quasiparticle spectra, our STS studies of the optimally doped Y-123 yield spatially homogeneous $\Delta_{SC}$ and less homogeneous $V_{CO}$ [30,35]. Similarly, our STS studies of the optimally doped La-112 also reveal highly homogeneous spatial distributions of $\Delta_{eff}$ [15]. These results are in stark contrast to the findings of a highly inhomogeneous effective quasiparticle excitation gap $\Delta_{eff}$ in Bi-2212 [31,37,55]. The strong inhomogeneity in $\Delta_{eff}$ of the Bi-2212 system may be attributed to its extreme two-dimensional nature so that disorder has more significant effect (such as pinning effect) on the low-energy quasiparticle excitations. If we assume that in Bi-2212 the inhomogeneous effective excitation gap $\Delta_{eff}$ is primarily attributed to inhomogeneous $V_{CO}$ and that the CO is associated with incommensurate CDW, we can use realistic energy histograms of $\Delta_{eff}(\delta)$ [31,55] to simulate spatial maps of $\Delta_{eff}$, LDOS and FT-LDOS that are consistent with empirical observation [55], as exemplified in Fig. 8 [58]. On the other hand, if we attribute the inhomogeneity to $\Delta_{SC}$, the corresponding LDOS would have been strongly varying near zero bias [58], which contradicts the experimental findings of highly homogeneous spectra near zero bias [31,37]. We further note that our simulations of the LDOS and FT-LDOS indicate the necessity of involving COs to achieve results consistent with experiments [56,58], because the assumption of pure SC as in the one-gap model cannot account for various spectral characteristics, including the periodic modulations in the LDOS and the *non-dispersive* wave-vectors along the $(\pi,0)$ and $(0,\pi)$ directions in the FT-LDOS.

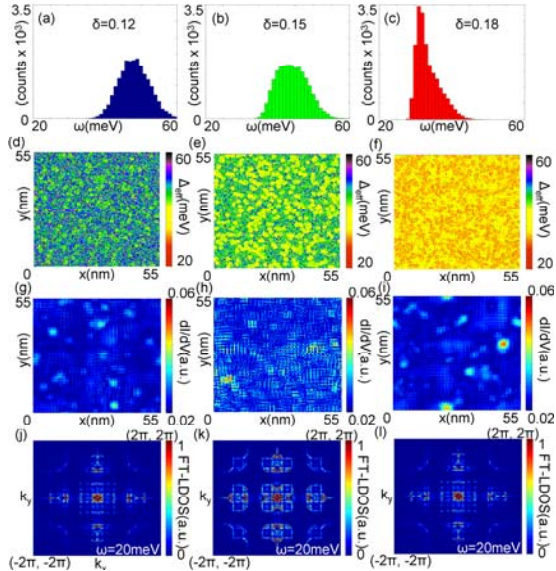

**Fig. 8** Simulations for doping-dependent quasiparticle spectral characteristics over a (55×55) nm² area based on the CO scenario [28,29,56,58] and the conjecture of coexisting homogeneous $\Delta_{SC}(\delta)$ and inhomogeneous $V_{CO}(\delta)$ associated with 50 impurities. Here the doping levels for the maps correspond to $\delta$ = 0.12 (1st column), 0.15 (2nd column), and 0.19 (3rd column), and the doping level determines the size of Fermi surface according to the realistic bandstructures of the cuprates [28,29]. 1st row **(a)** – **(c)**: energy histograms of $\Delta_{eff}(\delta)$; 2nd row **(d)** – **(f)**: $\Delta_{eff}(\delta)$ maps; 3rd row **(g)** – **(i)**: LDOS maps at $\omega$ = 20 meV; 4th row **(j)** – **(l)**: FT-LDOS maps at $\omega$ = 20 meV.

## 5 Discussion

While the CO scenario can provide a unified phenomenology for the quasiparticle excitation spectra in both electron- and hole-type cuprates, a number of issues remain unresolved. For instance, the microscopic mechanism for Cooper pairing in the presence of coexisting COs is still unknown. In addition, how different types of COs may emerge in the ground state of different cuprates is not fully understood. Moreover, a quantitative account for the quasiparticle LDOS, FT-LDOS and the spectral evolution with increasing magnetic field has yet to be developed theoretically. Finally, whether the presence of COs is helpful, irrelevant or harmful to the occurrence of high $T_c$ superconductivity in the cuprates awaits further investigation.

In addition to the existence of COs, it is worth commenting on possible variations in the pairing potential of the SC state. While the pairing symmetry of most cuprates is predominantly $d_{x^2-y^2}$, mixed pairing symmetries (such as $d_{x^2-y^2}$ + $s$) have been widely reported in a number of cuprates, including in the tunneling junction studies [35,59,60], phase sensitive measurements [61,62], microwave spectra [63], optical spectra [64], and $\mu$SR penetration depth measurements [65,66]. Moreover, the subdominant $s$-wave component appears to increase with increasing hole doping [35,64]. This finding is consistent with the notion that the $d_{x^2-y^2}$-wave pairing is more favorable when onsite Coulomb repulsion is significant near the Mott insulator limit, whereas the $s$-wave pairing component may become energetically preferred in the overdoped limit when cuprate superconductors become more like conventional superconductors. Regardless of the exact momentum dependence of the pairing potential, however, we emphasize that the low-energy excitation spectra of the cuprates cannot be accounted for by assuming a pure SC order in the ground state.

## 6 Conclusion

In conclusion, we have shown both experimental and theoretical evidences that the scenario of coexisting competing orders and $d_{x^2-y^2}$-wave superconductivity in the ground state of cuprate superconductors can provide a unified phenomenology that accounts for the unconventional and non-universal low-energy quasiparticle excitations among hole- and electron-type cuprates of varying doping levels. In particular, our vortex-state spatially resolved scanning tunneling spectroscopic (STS) studies of Y-123 have revealed various novel spectral characteristics, including two energy scales ($V_{CO}$ and $\Delta'$) other than the SC gap $\Delta_{SC}$ inside vortices and three accompanying energy-independent wave-vectors $\mathbf{Q}_{CDW}$, $\mathbf{Q}_{PDW}$ and $\mathbf{Q}_{SDW}$.

These results cannot be reconciled with theories assuming a pure SC order in the ground state. Similarly, our vortex-state spectroscopic studies of the electron-type cuprate La-112 also reveal an energy scale $V_{CO} < \Delta_{SC}$ inside vortices and a steady spectral shift from $\Delta_{SC}$ down to $V_{CO}$ with increasing magnetic fields. Moreover, we have shown that the presence of competing orders can account for the ARPES data that reveal the appearance (absence) of Fermi arcs in hole-type (electron-type) cuprates above $T_c$. The puzzling phenomena of strong spatial inhomogeneity in the quasiparticle tunneling spectra in Bi-2212 are also explained in the context of disorder-pinned competing orders so that the phenomena are not universal among different cuprates, as demonstrated by our own STS studies. Our investigations imply strong interplay between the pairing state and various collective low-energy bosonic excitations in the cuprates, thereby imposing important constraints on the microscopic descriptions for high-temperature superconductivity.

**Acknowledgements** This work was jointly supported by the Moore Foundation and the Kavli Foundation through the Kavli Nanoscience Institute at Caltech, and the NSF Grant DMR-0405088. The work at Sogang University was supported by the Center of Superconductivity from the program of Acceleration Research of MOST/KOSEF of Korea and Special fund of Sogang University. The authors thank Dr. A.I. Rykov for growing the YBa$_2$Cu$_3$O$_{7-\delta}$ single crystal used in this work and Professor S.A. Kivelson, Professor S.-C. Zhang, and Professor R. A. Klemm for useful discussions. ADB acknowledges the support of Intel Graduate Fellowship.